\documentstyle[12pt]{article}
\begin{document}
\centerline{\large \bf  Variation of the Energy Landscape}
\vskip 0.5cm
\centerline{\large \bf  of a Small Peptide under a Change}
\vskip 0.5cm
\centerline{\large \bf  from the ECEPP/2 Force Field to ECEPP/3}
\vskip 3.0cm
\centerline{Frank Eisenmenger$^{\dagger}$ and
Ulrich H.E.~Hansmann$^{\#}$
\footnote{New address: \em Institute for Molecular Science (IMS), Okazaki 444,
                 Japan}}
\vskip 1.5cm
\centerline{$^{\dagger}$\em Institute for Biochemistry, Medical Faculty
           (Charit{\`e})}
\centerline{\em of the Humboldt University Berlin, 10115 Berlin, Germany}
\vskip 0.5cm
\centerline{$^{\#}$\em Swiss Center for Scientific Computing (SCSC),}
\centerline{\em  Federal Institute of Technology (ETH)
                Z{\"u}rich, 8092 Z{\"u}rich, Switzerland}

\medbreak
\vskip 3.5cm
\centerline{\bf ABSTRACT}
The {\it multicanonical ansatz} is used to study variations in the energy
landscape of a small peptide, Met-enkephalin, under a change from the
ECEPP/2 force field to ECEPP/3. Local minima with energies up to $5~kcal/mol$
 higher
than the global minima are sampled and classified according to H-bridges
and backbone angles. The distribution and relative weight for various 
temperatures of the minima are calculated and compared for the two
force fields. We demonstrate that while there are small differences in 
the energy landscape  our  results at relevant temperatures are robust under
changes between ECEPP/2 to ECEPP/3.
\vfill
\newpage
\baselineskip=0.8cm
\section*{Introduction}

While there is considerable progress in numerical simulation of peptides and
proteins (for a recent review see, for instance, \cite{vas95}),
prediction of their low temperature conformations  solely from first
principles remains a formidable task.  First, the interactions between
the atoms in the molecules are described by  force fields which depend
in part on empirically determined parameters. It is still an open question which
of the many force fields (AMBER \cite{AMBER}, CHARMM \cite{CHARM}, ECEPP 
\cite{ecep75} - \cite{ecep92}, ...) is the optimal choice (or if they 
are sufficiently accurate at all).  
The other mayor problem is common to many important physical systems. 
At low temperatures the energy landscape of proteins is characterized by
a multitude 
of local minima separated by high energy barriers. In simulations  based on
classical  molecular dynamics or 
Monte Carlo techniques \cite{metro53}  these barriers can be 
seldom crossed. Hence, only small parts
of phase space are sampled (in a finite number of Monte Carlo sweeps) and
physical quantities cannot be calculated accurately. For this reason,
the ``native'' conformation of a protein (which should correspond to
the global minimum in {\it free} energy) is often identified with the
lowest {\it potential} energy configuration. In this approximation  entropic 
contributions are ignored, but it allows  application of
 optimization techniques
like simulated annealing \cite{SA}, genetic algorithms \cite{Fo} or
Monte Carlo with Minimization \cite{LS} to tackle the protein folding problem. 
However, this approximation may be
to crude.  Especially for smaller peptides entropic contributions are
expected to be important. 

It is not always clear whether the limited quality of numerical results
is due to insufficient simulation algorithms or inadequate force fields.
Recent progress in the development of Monte Carlo techniques may change this 
situation. Simulations in {\it generalized ensembles} promise a much better
sampling of the phase space. A numerical
comparison of some of these new algorithms can be found in Ref.~\cite{HO96b}.
One of its more  prominent exponents is  the so called {\it multicanonical} 
approach of Berg and co-workers \cite{berg91}  who propose a weighting scheme
which yield to 
a flat probability distribution in energy.  Hence, all energies have equal
weight and  an one-dimensional random walk in energy space is carried out 
(when simulated with local updates) which ensures that the simulation 
will not get trapped in any local energy  minimum. The Boltzmann
distributions may be obtained for a
given range of temperatures from one multicanonical simulation 
by re-weighting the states \cite{swen88}. 
The method was  exploited to model first-order phase transitions
 \cite{berg91,MU2} and spin-glass systems \cite{SG1,SG2,SG3}. 
The prediction of peptide and protein three-dimensional structures with
 multicanonical algorithms was first addressed 
 in Ref.~\cite{uli93} for Monte Carlo methods and in Ref.~\cite{hoe}
for molecular dynamics. 
 Subsequent works include the study of  coil--globular
transitions of a model protein \cite{HS}, helix-coil transitions of amino acid
homo--oligomers \cite{HO95} and the conformational sampling of a constrained
peptide \cite{KI95}.

In the current article we use the multicanonical 
technique to compare  distribution and relative weight of the 
local minima at room temperature and below for the closely related 
ECEPP/2 and ECEPP/3 force field in the case of the small pentapeptide 
Met-enkephalin. ECEPP/3 deviates from the previous ECEPP/2 
by slightly different parameters due to improvements in the experimental
measurements. We want to study the influence of such small changes on
the energy landscape  and our observed quantities. This should show us
how much numerical simulations are affected by experimental uncertainties
in the parameters of the  force fields.

\section*{Methods}

\subsection*{Multicanonical Algorithm}

Although the multicanonical approach is explained in detail elsewhere 
(see, for instance Refs.~\cite{SG1,uli94a}), we briefly
summarize the idea and implementation of the method for completeness.

Simulations in the canonical ensemble weight each configuration with
$w_{B}(E,\beta) =  e^{-\beta E}$
and yield the usual 
Boltzmann probability density distribution of energies:
\begin{equation}
P_B(E,T) \propto n(E) e^{-\beta E}
\end{equation}
where \quad $n(E)$ is the density of states with energy $E$, 
and \quad $\beta = \frac{1}{k_BT}$; with temperature $T$ and Boltzmann
constant $k_B$.  

On the other hand, in the multicanonical approach configurations with
energy $E$ are updated with a weight:
\begin{equation}
 w_{mu} (E)\propto n^{-1}(E) = e^{-S(E)}
\end{equation}
where $S(E)$ is the microcanonical entropy. A uniform distribution
of the energy 
\begin{equation}
 P_{mu}(E) = n(E)w(E) = {\rm const}
\end{equation}
is obtained from the simulation with these weight factors.
 In this way information  is collected over 
the whole energy range  and from a single simulation one can obtain the 
 canonical distribution   for a wide 
 range of temperatures by the re-weighting techniques:\cite{swen88}
 \begin{equation}
 P_B(T,E) \propto P_{mu} (E) w^{-1}_{mu} e^{-\beta E}~.\label{erw}
\end{equation}
This allows one to calculate any thermodynamic quantity ${\cal O}$ by
\begin{equation}
< {\cal O} >_T ~= \frac{\displaystyle{\int dE~ {\cal O} (E) P_B(T,E)}}
		       {\displaystyle{\int dE~ P_B(T,E)}}~.
\end{equation}

However, unlike in a canonical simulation the weights $w_{mu} (E)$ are not 
{\it a priori} known (in fact, knowledge of the exact weights is equivalent
with solving the system), and one needs its estimators for a numerical 
simulation.   One way to obtain these estimators is the following iterative
procedure.
 Starting with an initial guess $w^0_{mu}(E) = e^{-\beta_0 E}$,
 i.e., performing a canonical simulation at sufficient high temperature
 $T_0$,  improved estimators of the multicanonical
 weights are  
 calculated from histograms $P^{i-1}_{mu} (E)$ of preceding
 simulations by
\begin{equation}
w^i_{mu} (E) = \frac{w^{i-1}_{mu} (E)}{P^{i-1} (E)}.
\end{equation}
For details, see Ref.~\cite{HO95} and \cite{uli94a}. This method for
calculating multicanonical weights is by no means unique and while it
is quite general, it has the disadvantage that it requires a certain 
number of extra iterations which is not {\it a priori} known.

\subsection*{Force-Fields}

In ECEPP \cite{ecep75}--\cite{ecep92} 
 the potential energy function $E_{tot}$ is given by the sum of
the electrostatic term $E_{C}$, 12-6 Lennard-Jones term $E_{LJ}$, and
hydrogen-bond term $E_{HB}$ for all pairs of atoms in the peptide 
together with the torsion term $E_{tor}$ for all torsion angles:
\begin{eqnarray}
E_{tot} & = & E_{C} + E_{LJ} + E_{HB} + E_{tor},\\
E_{C}  & = & \sum_{(i,j)} \frac{332q_i q_j}{\epsilon r_{ij}},\\
E_{LJ} & = & \sum_{(i,j)} \left( \frac{A_{ij}}{r^{12}_{ij}}
                                - \frac{B_{ij}}{r^6_{ij}} \right),\\
E_{HB}  & = & \sum_{(i,j)} \left( \frac{C_{ij}}{r^{12}_{ij}}
                                - \frac{D_{ij}}{r^{10}_{ij}} \right),\\
E_{tor}& = & \sum_l U_l \left( 1 \pm \cos (n_l \chi_l ) \right).
\end{eqnarray}
Here, $r_{ij}$ is the distance between the atoms $i$ and $j$, and 
$\chi_l$ is 
the torsion angle for the chemical bond $l$. 
In ECEPP bond lengths and bond angles (which are hard degrees of freedom) are
 fixed at experimental values and no out-of-plane deformation of peptide
bonds is allowed leaving the dihedral angles $\phi,\psi,\omega $ and $\chi$
as independent variables. The various parameters ($q_i,A_{ij},B_{ij},C_{ij},
D_{ij},U_l$, and $n_l$) for the energy function were 
determined by a combination of {\it a priori} calculations and minimization
of the potential energies of the crystal lattices of single amino acids.
In the light of more recent experimental findings the standard geometry 
and some  energy parameters for prolyl and hydroxyprolyl have been updated. 
Together with a re-calculation of partial atomic charges of mainchain 
atoms this led to a revision of the ECEPP/2 parameter set to ECEPP/3.
In the original formulation of the ECEPP/2 parameter set, the net charges 
of both the N- and the C-terminal residues, added up to the desired net 
charge of the molecule. Now each terminal amino acid residue carries a
total charge of 0/+1/-1 at the N- or C-terminus, respectively. The 
correction avoids possible artifacts for systems with terminal groups of
different types (e.g. one terminus  charged, the other uncharged).

\section*{Simulation and Technical Details}

To investigate changes in the energy landscape under small variations
of the force field 
we have studied one of the simplest peptides,
Met-enkephalin which has the amino acid sequence Tyr-Gly-Gly-Phe-Met.  
This peptide is convenient for our purpose, since for the potential
energy function ECEPP/2 the lowest energy conformation 
is known \cite{LS,Yuko1,Hagai} and local minima with energies not much higher  
than the global minimum were sampled and classified for this molecule 
by Braun and co-workers.\cite{Braun} We compare our results with theirs
and calculate in addition the relative weight of the local minima at
various temperatures which is not possible with the method used by
Braun {\it et al.}

For our simulations the
backbone was terminated by a neutral --NH$_2$ group at the N-terminus
and a neutral~ --COOH group at the C-terminus as in the previous works of
Met-enkephalin. For Met-enkephalin the two versions of ECEPP differ in the
point charges for terminal groups which are listed in Tab.~1.
The peptide-bond
dihedral angles $\omega$ were fixed at the value 180$^\circ$, leaving 
 19 dihedral angles  as independent variables.\footnote{We fixed these
hard degrees of freedom to reduce the number of variables. It would be
more rigorous to allow for bending of the peptide angles. Similarly, one
could argue that the fixed geometry of ECEPP should be replaced by a 
flexible one (as in other force fields), for instance by including 
Fixman potentials. However, in this  article we are mostly interested  in
comparing two variants of a force field, not in  reproducing
experimentally found structures by numerical simulation, which allow us
to choose a less rigorous approach.}
Interactions of the molecule with
solvent molecules were neglected
and the dielectric constant $\epsilon$ was set equal to 2.
We used the program  SMC \cite{eisen}  which was modified to accommodate
the multicanonical ensemble.

The multicanonical
weight factors were determined separately for both ECEPP/2 and ECEPP/3 
by the iterative procedure described
above. We needed  100,000 sweeps 
for their calculation. 
One MC sweep updates every 
dihedral angle  of the 
molecule once. For each update of an angle a new randomly chosen
value out of the
interval $[-\pi,\pi]$ was proposed and the new configuration then 
accepted or rejected by the Metropolis criteria. 
 All thermodynamic quantities were   calculated  for ECEPP/2 and 
ECEPP/3 separately from
a production run of 200,000 MC sweeps following additional 10,000
sweeps for equilibration each. The energy after each sweep was stored
for further analysis. 
In all cases, each simulation started from a completely random initial
conformation (``Hot Start''), but in the case of ECEPP/2 
we also checked our results for 
shorter runs against these  with initial configuration of the known
groundstate (``Cold Start''),
and we found that the results are in agreement with
those from random initial conformations.  
This suggests that thermal
equilibrium has been attained in our simulations. This conjecture is
further supported by Fig.~1,
in which we show the timeseries of the energy for both production runs.
In the course of the simulation,  a 1d random walk in the energy 
between low energy states and high energy states is performed
as one would expect for local updates 
from the definition of the multicanonical ensemble. It is evident
that low energy states which are separated in the timeseries by
high energy states (which correspond to high temperature states)
are uncorrelated. The number of such ``tunnel'' events is therefore
a lower limit for the number of independent low energy states visited
in the simulation.

In order to investigate the energy landscape of our peptide we not only
stored the energy of the actual conformation after each
MC sweep, but also minimized a copy of this conformation if its
energy was less than a  certain limit, $0~kcal/mol$. While this limit is 
somehow
arbitrary, our experience shows that it allows a rough distinction between
a ``low temperature region'' and a ``high temperature region'' (it is the
expectation value of Energy at $T=300 K$ where the specific heat has its
maximum). From the minimization process
we receive a local minimum conformation. Configurations which yield
the same local minima belong to the same valley in the energy landscape
and are therefore related to each other. Since we are mainly interested in
local minima not too far away from the global minimum and to save 
disc space,  we introduced a second
limit. Only if the energy of the minimized configuration was 
below $-6~kcal/mol$ ($\approx 5~kcal/mol$ above the
global minimum) it was stored for further analysis. In Tab.~2 we
summarize the absolute number of local minima sampled in this way
for  both variants of the ECEPP force field.

While the set of configurations we received from our production runs are all
local minima,  due to limitations of the minimizer (a Newton-Raphson
variant) they are not necessarily
the lowest configurations associated with a certain valley in the energy
landscape. Each valley is in itself rough and consists of  subvalleys
separated by energy barriers. Since we are not interested in the 
microscopic details of the energy landscape we further clustered our local 
minima and identified each cluster by its member with the lowest energy.
In this way, only a small number (less than 5) of groups remains at the
end of our clustering  
by which the overwhelming number of local minima is classified.

We describe now briefly our clustering technique. In a first step we
sorted all minimized conformers according to their energy. In a second step, we
looked for identical conformers. Two conformers were regarded as identical
if they differ in no angle more than $1^{\circ}$ (taking symmetries in account)
and in their energies less than $0.0001~kcal/mol$. For further clustering, we 
kept only the non-identical conformations. Their number which is also
shown in Tab.~2 is  our estimate for the total
number of local minima with energy below $-6~kcal/mol$. 
It is interesting to observe that the
number of non-identical conformers sampled in the multicanonical runs is 
similar for each choice of force field while the absolute number of sampled 
local minima varies much more (see Tab.~2). 
In a third step, using all backbone angles,  the conformer (from our remaining 
set of non-identical local minima conformations) with the lowest energy is
compared to all other with higher energies, according to differences
in their backbone dihedrals. The conformers having all their angles
within $10 ^{\circ}$ in comparison with the first one, with respect to
symmetry, are put into the same cluster, which is represented by the first
conformer. This procedure is repeated for the next conformer with higher energy,
which was not counted in the first cluster by comparing its angles with
all the other remaining "free" conformers.  In this way,  we are left with
160 clusters for ECEPP/2 and 571 clusters for ECEPP/3. In the final step  
these clusters were classified into few groups using similarities
in hydrogen bonding and the significant dihedrals, necessary to preserve
 the  characteristic hydrogen bonds. The final groups 
which should represent the main valleys in the energy landscape of our
peptide are  characterized by their lowest-energy conformer.

In Fig.~2a and 2b we display how in the course of our simulations
different valleys of the energy landscape (characterized by its 
corresponding group number) are visited. We show only that times when the
simulation visit one of the 4 (3)  most important clusters for
ECEPP/2 (ECEPP/3)  and no plot symbol means that the simulation did not
visit one of these valleys. As one can see there are different slices
in which the multicanonical simulation moved only inside of a   valley
separated by jumps between different valleys. Again the number of
these slices will give a lower limit on the number of independent
low energy states visited in the simulation. 

\section*{Results and Discussion}

Using the techniques described above we were able for each choice of the force 
field to classify the large
number of local minimum conformations sampled in our multicanonical runs
into few groups which represent more than 80\% of
the sampled local minimum conformations and  which correspond to the important 
valleys in  the energy landscape of Met-enkephalin. 
In Fig.~3a and Fig.~3b 
the dominant groups for both variants of the ECEPP force field are shown. 
In addition, we present  in Tab.~3 the dihedral angles for the 
global minimum conformations obtained by our method. These structures are 
also shown in Fig.~4a and Fig.~4b.

The global minimum for ECEPP/3 has characteristic hydrogen bond of the NH-group
of Phe(4) with the carbonyl group of Try(1) resulting in a $\beta$ -bend  between
 the first and fourth residue. To preserve this hydrogen bond only the
backbone angles of Gly(2) and Gly(3) have to be conserved and therefore 
one finds a large number of variants differing from the global minimum
stucture in the other backbone and sidechain dihedrals in this group
(which we call group ``1''). The second group (``2'') is characterized by the
local minimum configuration with  the second lowest energy in the ECEPP/3 force
field. This configuration has two possible hydrogen bonds between N-H and
C=O groups of the Gly(2) and Met(5) residue forming a II'-type $\beta$-turn 
between the second and last residue. Since more backbone dihedrals have to be
conserved to preserve the two hydrogen bonds the configurations of this group
vary much less than in the case of the previous group. Despite their
differences (see Figs.~3 and 4)  the lowest
energy exponent of both groups differs by only  $0.04~kcal/mol$ in potential
energy, so that one can regard the global minimum of Met-enkephalin in
the ECEPP/3 force field as degenerated.  However, the electrostatic energy 
for the ECEPP/3-minimum is raised substantially if ECEPP/2 charges are used 
for this structure. The total potential energy 
increases from $E_{tot} = -10.84~kcal/mol$ (ECEPP/3) to $E_{tot} = -9.46
 ~kcal/mol$ (ECEPP2) while the energy of the other configuration changes only 
from $E_{tot} = -10.81~kcal/mol$ (ECEPP/3) to $E_{tot}= -10.72~kcal/mol$ (ECEPP/2),  
which is the global minimum of Met-enkephalin in the ECEPP/2 force field
\cite{Yuko1,Hagai}. 
Configurations of both group ``1'' and ``2'' were also found in our ECEPP/2 
simulation.

We remark that  our peptide seems to have  a much richer spectrum of low energy 
local minimum configurations  in the ECEPP/3 force field than in the case of 
ECEPP/2, but show in Fig.~3b only one more (the other contribute with less
than 1\% each) group. Conformers
of this group (``H'') are characterized by a hydrogen bond between 
C=O group of the 
Tyr(1) and the N-H group of Met(5) residues. Together with the backbone
dihedrals of Gly(2), Gly(3) and Phe(4) this hydrogen bonding resembles that 
of an $\alpha$-helix. Its lowest energy exponent has a potential energy of
$-10.56~kcal/mol$ which is only $0.25~kcal/mol$ higher than the groundstate. 
No such conformation was found in the ECEPP/2 simulation.
In the case of ECEPP/2, two more groups appeared with a frequency of more
than 1\% and are shown in Fig.~3a. Both are II' type $\beta$-turns. Group ``a''
is characterized by single hydrogen bond  between  Gly(2) and Met(5).
The lowest energy conformation in this group has a potential energy of
$-9.74~kcal/mol$ which make this group to the one with the second lowest
potential energy, separated by $\approx 1~kcal/mol$ from the groundstate. Group
``b'' has its hydrogen bond between    Tyr(1) and Phe(4), its lowest energy
exponent has a potential energy of $-9.52~kcal/mol$). Both groups 
were also found in the ECEPP/3 simulation, but with frequencies less than 1\%
of the non-identical conformers. 

For ECEPP/2 our local minimum conformations
resemble those  found in previous work \cite{Braun} where in difference to
the present work 
the peptide angles were released.  Fixing these angles  therefore seems to be
a reasonable approximation. But while our classification of low energy
local minima  can also be obtained with other methods,  
the multicanonical approach allows in addition the estimation of 
the relative weight of these conformers as a function
of temperature, i.e.  entropic contributions can be considered.
As an example we show in Fig.~5a and 5b the relative weight 
of the dominant groups of conformation for both ECEPP/2 and  ECEPP/3.
Note that around $300 K$ the percentage of configurations which are in
none of our groups (and hence do not belong to the important valleys)
decreases dramatically for both force fields. At high temperatures coil
structures dominate which show large flexibility and and therefore  a broad
energy spectrum. On the other hand, at low temperatures ordered structures
are expected which are each confined to  one of the valleys in the
energy landscape.  
In the case of the ECEPP/2 simulation the percentage of configurations of type
`1' (the ECEPP/3 groundstate) never exceeded 4\%.
 Configurations of type `2' (the groundstate for ECEPP/2)
are found with much higher probability, but at room temperature they
contribute to only about 30 \% of the conformers which confirms older
work \cite{uli93}. For ECEPP/3 we find that 
the probability of configurations of type `2' is comparable to that of the
ECEPP/2 simulation. At room temperature they contribute again to around
30\% of the conformers. Conformations of type `1' appear with much
higher probability than in the ECEPP/2 simulation, but that of type `2' are
still dominating in the range of shown temperatures. At relevant
temperatures the probability for finding the ECEPP/2 groundstate (type `2')
 is twice as
high as that of finding the ECEPP/3 groundstate (type `1').
Of course we expect that at
 $T=0~K$  we will  find only the ECEPP/3 groundstate (type `1'), however
the energy difference  is much too small (about $0.04~kcal/mol$)
 to favor them at finite temperatures against type `2' structures.
This proofs that at least for small peptides entropic
contributions cannot be neglected. Any pure optimization method
(in the potential energy) may lead to configurations which are of only
limited significance at relevant temperatures.

In the final plots we like to show how the variations in the force field
affects two important thermodynamic quantities. Fig.~6 displays
the average energy $<E>$ as a function of temperature.
Again, we observe only little differences between the ECEPP/2 and ECEPP/3
force fields for the physical relevant temperature range. Below $400$ K
these differences are within the errorbars, while for higher temperatures
the ECEPP/3 energies are systematically lower by a small amount.
Similar results were found for
the specific heat $C(T)$ as function of temperature  shown in Fig.~7. 
Here we define
\begin{equation}
  C(\hat{\beta})  = {\hat{\beta}}^2 \frac{<E^2> - <E>^2}{5}~.
\end{equation}
 Neither the position of its maximum nor its shape or heights
vary much indicating that the transition between ordered and disordered
states is of the same kind for both variants of the force field. Note that
the position of the peak in specific heat corresponds to the increase 
of structures which belong to one of the main groups and the decrease of
configurations which do not belong to any of the significant energy valleys.

\section*{Conclusion}
We used the multicanonical approach to study variations of the energy
landscape of Met-enkephalin under a change from the ECEPP/2 force field to
ECEPP/3. In both cases, the 
low-energy local minima were sampled and classified into
a small number of groups which correspond to essential valleys of the
energy landscape.  We studied their distribution and relative
weight at various temperatures. While the energy landscape differs in detail
and specially  the global minimum is not the same, 
 our physical results are little affected
by changes between the two force fields. 
Our analysis demonstrates that at least for small peptides  it is not
sufficient to search only for the global minimum in {\it potential}
energy since this conformation may be of limited significance at room 
temperature.
Our work corroborates the well known fact that small flexible molecules
exist at room temperature in an ensemble of low energy conformations 
(see for instance Ref.~\cite{vas95} and work cited therein) whose relative
weight has to be determined. Finally, our results show that 
numerical simulations of peptides are stable under small changes in the
utilized force fields. Especially thermodynamic quantities are little changed.
This is an important observation since all
force fields rely on experimentally determined parameters and are therefore
known only within certain errorbars.

\vspace{1.5cm}
\noindent
{\large \bf Acknowledgements}:\\ 

\noindent
This work was supported, in part, by the Schweizerische Nationalfonds 
(Grant 20-40'838.94) and  by the U.S. Department of Energy  
(contract  DE-FC05-85ER250000).
The simulations were performed on the cluster of  RISC 
workstations at SCRI, The Florida State University, 
Tallahassee, USA, and the Institute for Biochemistry
at the Humboldt University, Berlin, Germany.

\newpage
\noindent
{\bf \Large TABLE CAPTIONS:}\\
TAB.~1: {Point Charges for Terminal Groups of Met-Enkephalin.}\\
TAB.~2: {Number of conformers and non-identical conformations
         collected after minimization for  both choices of force fields.\\
TAB.~3: {Dihedral angles of our estimates of the global minima for ECEPP/2 
         and ECEPP/3. The peptide bond angles were fixed to 
         $\omega = 180^{\circ}$} in the simulations.\\

\newpage
{\Large Table 1:}\\ \\
\begin{center}
\begin{tabular}{l r r}
            Atom &     ECEPP/2 &   ECEPP/3\\ \hline\hline
            N     &    -0.356  &   -0.332 \\
            H1,H2 &     0.176  &    0.076 \\ \hline
            C     &     0.450  &    0.517 \\
            O     &    -0.384  &   -0.351 \\
            O     &    -0.380  &   -0.334 \\
            H     &     0.204  &    0.235 \\
\end{tabular}
\end{center}

\newpage
{\Large Table 2:}\\ \\
\begin{center}
\begin{tabular}{ | l | l  |  l | } \hline
        &  ECEPP/2 &   ECEPP/3  \\ \hline
Minimized conformers $E < -6 ~ kcal/mol$ &  62142 &   109863 
\\
  \hline
Non-identical minimized conformers                     &         &      \\
($\ge 1$ dihedral differs by $>1^{\circ}$ ) & 1677     &  1574   \\
  \hline
\end{tabular}
\end{center}

\newpage
{\Large Table 3:}\\ \\
\begin{center}
\begin{tabular}{ | l c | c   | c   | } \hline
  &      &  ECEPP/2 &   ECEPP/3  \\ \hline
  \hline

 1 ~ Tyr &  $\phi$    & -86.3  & -162.7 \\
         &  $\psi$    &  153.7 & -41.7  \\
         &  $\omega$  & 180.0  & 180.0  \\
         &  $\chi_1$  & -179.8 & -174.2 \\
         &  $\chi_2$  & -111.4 & -85.2  \\
         &  $\chi_6$  & 145.3  & 2.8    \\ \hline
 2 ~ Gly &  $\phi$    & -161.5 & 65.8   \\
         &  $\psi$    & 71.1   & -87.0  \\
         &  $\omega$  & 180.0  & 180.0  \\ \hline
 3 ~ Gly &  $\phi$    & 64.1   & -157.3 \\
         &  $\psi$    & -93.5  & 34.9   \\
         &  $\omega$  & 180.0  & 180.0  \\ \hline
 4 ~ Phe &  $\phi$    & -81.7  & -158.8 \\
         &  $\psi$    & -29.2  & 159.5  \\
         &  $\omega$  & 180.0  & 180.0  \\
         &  $\chi_1$  & 179.8  & 52.4   \\
         &  $\chi_2$  & -100.0 & -96.0  \\  \hline
 5 ~ Met &  $\phi$    & -80.7  & -82.4  \\
         &  $\psi$    & 143.5  & 134.1  \\
         &  $\omega$  & 180.0  & 180.0  \\
         &  $\chi_1$  & -65.1  & -66.1  \\
         &  $\chi_2$  & -179.2 & -179.6 \\
         &  $\chi_3$  & -179.3 & -179.9 \\
         &  $\chi_4$  & 60.1   & 60.1   \\ \hline
 $E_{tot}$(ECEEP/2) [$kcal/mol$] &  & -10.72   & -9.46  \\
 $E_{tot}$(ECEPP/3) [$kcal/mol$] &  & -10.81   & -10.85 \\ \hline
\end{tabular}
\end{center}

\newpage
\noindent
{\bf \Large FIGURE CAPTIONS:}\\
FIG.~1:   Timeseries of potential energy, obtained from a multicanonical
          simulation of 200,000 MC sweeps, using  a) the ECEPP/2 and 
          b) the ECEPP/3 force field.\\

\noindent
FIG.~2:   Timeseries of clusters, obtained from a multicanonical
          simulation of 200,000 MC sweeps, using a) the ECEPP/2 and
          b) the ECEPP/3 force field.\\

\noindent
FIG.~3:   Most frequent low energy conformers  for a) ECEPP/2 and b) ECEPP/3.
          Shown are also their Zimmermann
          codes, the frequency with which they appear in the multicanonical
          simulation and the potential energy of their lowest energy
          exponent $E_{min}$.\\

\noindent
FIG.~4:   Groundstate configurations of a) ECEPP/2 and b) ECEPP/3. The
          plots were drawn using RASMOL.\\

\noindent
FIG.~5:   Relative weight of the  important groups of configurations
          in the case of a) ECEPP/2 and b) ECEPP/3.  We have sum up
          all local minima  which do not belong to one of the 
          mayor groups in ``Other''. Likewise  ``No Minimum'' contains
          all configurations which could not be minimized to a local
          minimum with potential energy below $-6~kcal/mol$ (see text).\\

\noindent
FIG.~6:   Average potential energy as a function of temperature for  both
          choices of force-fields. \\

\noindent
FIG.~7:   Average of specific heat as a function of temperature for  both
          choices of force-fields. \\
\end{document}